\documentclass[reprint,amsmath,amssymb,aip]{revtex4-2}

\usepackage[dvipsnames]{xcolor}
\usepackage{soul}
\sethlcolor{yellow}

\usepackage{lmodern}
\usepackage[english]{babel}
\usepackage{graphicx}
\usepackage{iftex} \ifpdftex\usepackage[T1]{fontenc}\fi
\usepackage{nicefrac}

\begin{document}

\onecolumngrid
This article may be downloaded for personal use only. Any other use requires prior permission of the authors and AIP Publishing. This article appeared in Armand Rykner, François Ladieu, Marceau Hénot; Fast temperature up steps as a test of the Tool–Narayanaswamy formalism. J. Chem. Phys. 29 September 2026; 165, 124514. and may be found at \url{https://doi.org/10.1063/5.0352723}.


\title{Fast temperature up steps as a test of the Tool-Narayanaswamy formalism}

\author{Armand Rykner}
\author{François Ladieu}
\author{Marceau Hénot}
\email{marceau.henot@cea.fr}
\affiliation{SPEC, CEA, CNRS, Université Paris Saclay,\\91191 Cedex Gif-sur-Yvette, France}


\begin{abstract}
	We investigated the aging dynamics of a glass-forming liquid triethyl-2-acetylcitrate (TEAC), following fast temperature up steps with amplitudes ranging from 0.3 to 13.6 K. The initial states were either at equilibrium or prepared at increasing levels of out-of-equilibrium through a prior down step experiment. Our goal was to test the predictive power of the Tool-Narayanaswamy (TN) formalism which assumes that the non-linear re-equilibration of a liquid can be linked to its linear response to a small perturbation. We determined the TN parameters for steps with small to moderate amplitude ($\leq 3.3$~K) and crucially took advantage of down step aging experiments below the glass transition temperature to constrain the determination of the equilibrium relaxation time. We tested the TN predictions and found very good agreement for differences in fictive temperature characterizing the distance from equilibrium as high as 10~K. For larger steps, however, the prediction progressively fails to capture the aging dynamics. This likely indicates that the re-equilibration mechanism is no longer related to the equilibrium dynamics. Finally, we discuss the possibility of obtaining a general criterion for the limit of validity of the TN formalism, which we compare to other systems from the literature.
\end{abstract}

\maketitle

\begin{figure*}
	\centering
	\includegraphics[width=\textwidth]{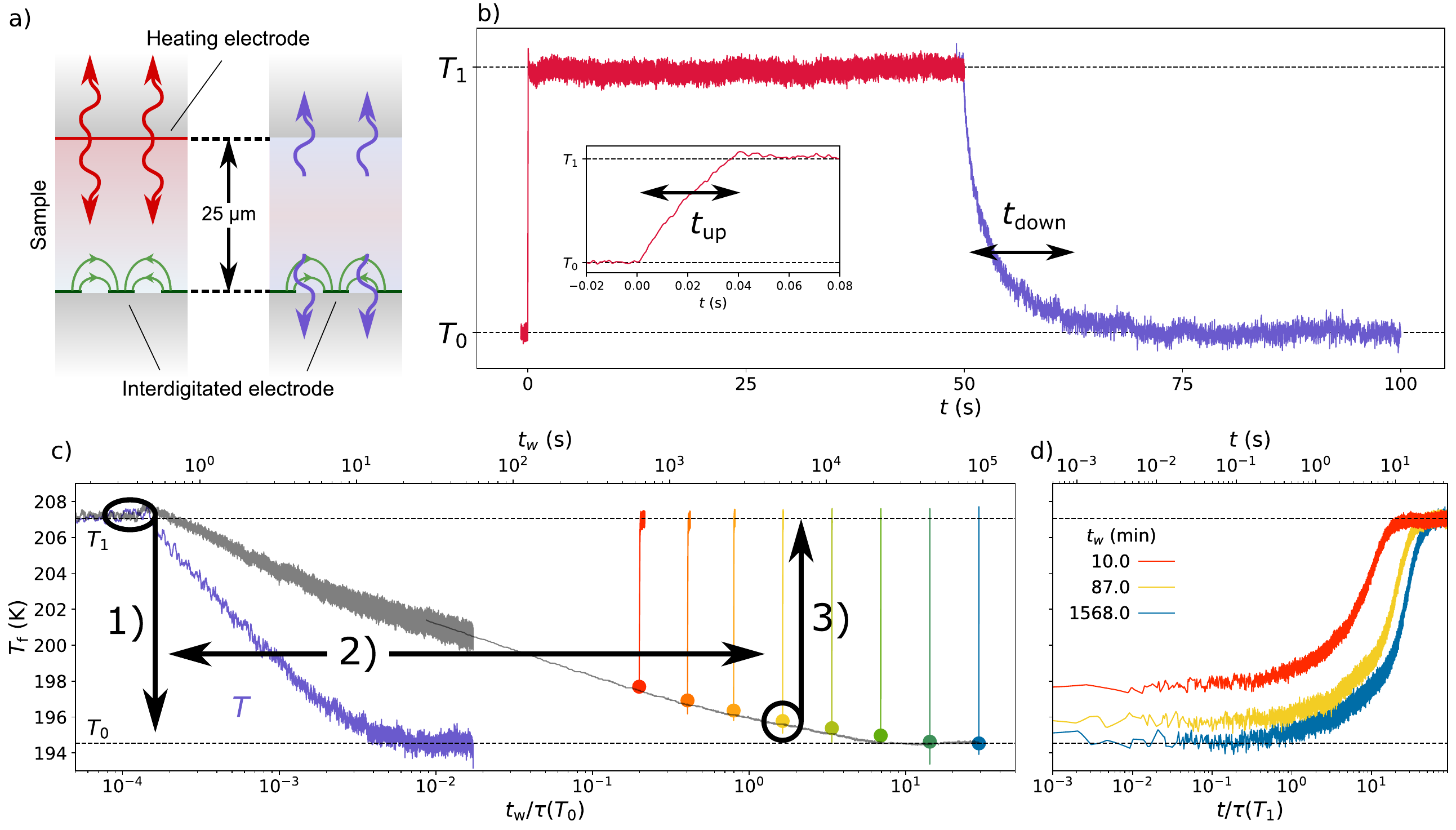}
	\caption{(a) Drawings of the cross-section of the experimental setup depicting heating and cooling processes. (b) Temperature profile of the sample when applying up step (in red, $t_\mathrm{up}=40$~ms) and down step (in blue, $t_\mathrm{down}\sim12$~s). (c) Representation of the complete experimental process: 1) We start from an equilibrium state at $T_1$ and apply a down step to $T_0$. 2) We wait for a time $t_\mathrm{w} \leq t_\mathrm{eq} \sim 10 \tau_\alpha$ ($t_\mathrm{eq}$ being the time required for it to reach its new equilibrium state at $T_0$) during which the sample slowly relaxes. 3) We apply an up step back to $T_1$ and measure its re-equilibration dynamics. (d) Re-equilibration curves for up step from 194.6 to 207.1~K from three different waiting times. These curves correspond to the ones of same color in (c).}
	\label{fig:Method}
\end{figure*}

\section{Introduction}\label{sec:level1}
Upon cooling, the molecular relaxation time of supercooled liquids displays a dramatic increase. This leads ultimately to a glass transition at temperature $T_\mathrm{g}$ where the equilibration time becomes so large that the system get stuck in an out-of-equilibrium state\cite{angell2000relaxation}. Close to $T_\mathrm{g}$, dynamics go through an exponential slowdown called aging, and full equilibration can only be observed with patient monitoring. A popular type of aging experiment is the temperature step (or jump) where the re-equilibration of the system is monitored following a change in temperature~\cite{Kovacs1963, Hecksher2010, Riechers2022, di2023physical, Henot2024, lancelotti2024kinetics}. Because the relaxation time is highly temperature dependent, a step with only a few Kelvin in amplitude $\Delta T$ leads to a non-linear response in the sense that the dynamics is strongly dependent on $\Delta T$. A consequence is the so-called asymmetry of approach to equilibrium~\cite{lillie1933viscosity, Kovacs1963, mckenna201750th}: a down step is faster than an up step to the same final temperature.

Since the combined work of Tool~\cite{Tool1946} and Narayanaswamy~\cite{Narayanaswamy1971}, the TN formalism has become the standard framework for describing the nonlinear relaxation dynamics of glasses near $T_\mathrm{g}$~\cite{Moynihan1976, Scherer1986, Hodge1994, Hecksher2015, malek2024distinguish, moch2024nonlinear}. By introducing a material time $\xi$, a transformed timescale that integrates the temperature and structure dependent relaxation rate, the formalism reduces the inherently nonlinear evolution in real time $t$ to a linear response in $\xi$, enabling quantitative predictions of equilibrium recovery and aging~\cite{Riechers2022}. Beyond its success as predictive tool, the validity of the material time approach was assessed both experimentally~\cite{Bohmer2024} and numerically~\cite{Douglass2022, amari2026large} for moderate temperature steps. However, cases in which this formalism fails have since been identified. Most notably in the case of ultra-stable glasses: very low-enthalpy glasses prepared by physical vapor deposition~\cite{Swallen2007, Ediger2017}. Upon heating, these glasses transform via heterogeneous mechanisms, such as front-mediated~\cite{swallen2009stable, sepulveda2014role, RafolsRibe2017} or nucleation-and-growth processes~\cite{Herrero2023, RuizRuiz2023, VilaCosta2023, tracy2024initial}. This heterogeneous mechanism is not unique to ultra-stable glasses and was also observed in standard glasses using high speed calorimetry technics~\cite{VilaCosta2023, Kaur2023}. The TN description is compatible with the existence of nanoscale dynamical heterogeneities, as observed at equilibrium~\mbox{\cite{vidal2000direct}}. But the whole distribution of relaxation times is assumed to age at the same rate given by the material time. The mean relaxation time is thus assumed spatially uniform and this approach cannot account for micrometer-scale heterogeneities during aging. This raises the question of how far the TN formalism validity range can be extended and whether it reflects a genuine microscopic mechanism or rather an effective description valid in a restricted regime of glassy relaxation.

In the present work, we studied the re-equilibration dynamics of triethyl-2-acetylcitrate (TEAC) following fast temperature up steps of amplitude ranging from 0.3 to 13.6~K starting from either an equilibrium or an out-of-equilibrium state. We used the smaller amplitude steps ($\Delta T \leq 3.3$~K) to determine the TN model parameters. We also relied on the results of down step aging experiments to constrain the temperature dependence of $\tau_\alpha(T)$ below $T_\mathrm{g}$. We then tested the TN predictions for the large up steps and we distinguished the agreement regarding the re-equilibration time and the quality of the prediction for the full re-equilibration dynamics. While the former was well predicted on the whole range studied, we identified a limit for the validity of the TN for the latter. We discuss this limit in the context of results reported in the literature for other systems.

\section{Method}\label{sec:level2}

\subsection{Experimental setup}\label{subsec:level2a}
Our measurements were performed on a triethyl-2-acetylcitrate sample (Sigma-Aldrich, 99\%, $T_\mathrm{g}=199$~K), through an experimental setup previously described in ref.~\cite{Henot2024} and which we briefly explain here. A thin liquid film was confined between two glass electrodes separated by a 25~$\mu$m thick kapton spacer. The top electrode was covered with a thin resistive indium tin oxyde layer that is used to quickly heat the sample. On the bottom electrode, we lithographed interdigitated electrodes (IDE) made of 240 pairs of copper lines. Their width and the gap between them are both equal to 6~$\mu$m. The IDE capacitance is related to the dielectric permittivity of the substrate $\epsilon_\mathrm{s}$ and of the liquid $\epsilon^\star$ within the first few micrometers of the sample next to the surface, by $C^\star=C_0(\epsilon_\mathrm{s} + \epsilon^\star)$, where $C_0 = 5$~pF. We measured the complex capacitance $C^*$ at frequency $f=1120$~Hz of the IDE using a resistor ($R=100$~k$\Omega$) connected in series with it, through either an RTB2004 oscilloscope, or an SR830 lock-in amplifier. With the former we extracted one datapoint per period.

Up-steps from $T_0$ to $T_1$ were performed by applying a $t_\mathrm{up}=40$~ms continuous current pulse through the resistive electrode, which rapidly heats the entire sample due to its thinness (see Fig.~\ref{fig:Method}a). This was followed by a sequence of short pulses~\cite{Henot2023,Henot2024} to maintain the sample temperature around $T_1$ within 2\% of $\Delta T$ for approximately 50~s. This process induces an temperature inhomogeneity within the sample, which we quantified in the Supporting Information. This mainly affects the first 100 ms of the measurement. After that, it causes only minute uncertainties within the data, smaller than the markers used in all figures. When the current is turned off, the small sample thickness leads to a temperature down step through thermal diffusion (see Fig.~\ref{fig:Method}a) with a characteristic time scale $t_\mathrm{down}\approx 12$~s. Both types of step were characterized (see Fig.~\ref{fig:Method}b) by operating at a temperature $T_0=185$~K, small enough compared to $T_\mathrm{g}$ so that no relaxation occurs on the measurement time window. The dielectric permittivity of the liquid $\epsilon^\star$ can thus be used as a thermometer. Assuming that the thermal response of the system remains in the linear regime, the temperature profile $T(t)$ for any other step can be deduced by re-scaling it. Typical values of $\Delta T$ achievable with our setup range from 0.25 to 18~K. In order to obtain a usable signal, step temperature experiments were averaged on up to 350 independent realizations for $\Delta T = 0.3$~K and more typically 20 realizations for $\Delta T = 4$~K. Experiments involving down steps to $T_0 = 196$~K were repeated 9 times while those at $T\leq 194.6$~K where performed only once. The starting temperature $T_0$ for each experiment is chosen to ensure that the steps are ideal, meaning that no relaxation occurs during the heating phase or, in other words, that the step is instantaneous from the perspective of the system.

\subsection{Data processing}\label{subsec:level2b}
The dielectric permittivity $C^*(t)$ captures both the temperature evolution of the sample and its re-equilibration dynamics. In order to separate these two contributions, we converted the real part of the capacitance $C^\prime$ into a fictive temperature $T_\mathrm{f}$ which characterizes the distance from equilibrium $T-T_\mathrm{f}$ of the system in temperature units~\cite{mauro2009fictive}. This was done in two steps~\cite{Henot2024}: we first converted $C^\prime(t)$ into a temperature $T_\mathrm{from~C^\prime}(t)$ using an equilibrium calibration $C_\mathrm{eq}^\prime(T)$. This allows one to take into account the non-linear relationship between the dielectric permittivity and the temperature. Second, we decomposed $T_\mathrm{from~C^\prime}$ into two contributions accounting for the phonon temperature $T$ and for the fictive temperature $T_\mathrm{f}$.
\begin{equation}
	\Delta T_\mathrm{from\,C^\prime}(t) = \kappa \Delta T(t) + (1-\kappa) \Delta T_\mathrm{f}(t)
	\label{eq:TfromC}
\end{equation}
where the parameter $\kappa$ is determined for each experiment in order to insure that $T_\mathrm{f}$ remains continuous during an ideal temperature step (see Supporting Information).

\section{Results}\label{sec:level3}

\subsection{Protocol}\label{subsec:level3a}
\begin{figure*}
	\centering
	\includegraphics[width=\textwidth]{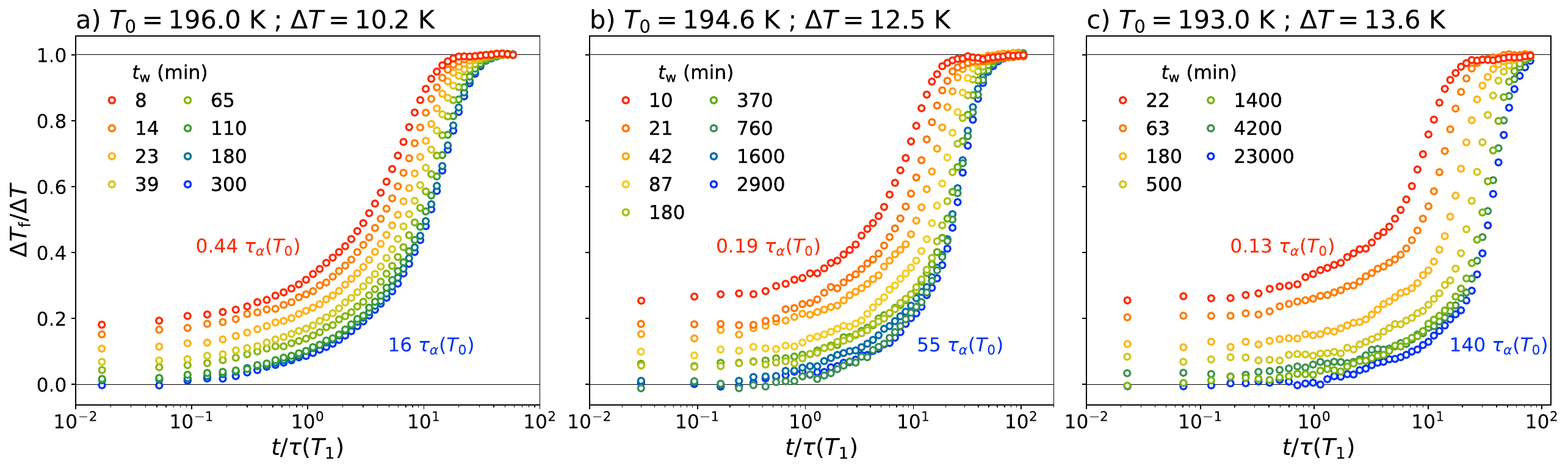}
	\caption{Normalized relaxation curves for up steps from out-of-equilibrium states produced through the protocol described in Fig.~\ref{fig:Method}. (a), (b) and (c) correspond to three different pairs of $T_0$ and $\Delta T$, and color encodes the waiting time, from red for small waiting times (i.e., far from equilibrium initial states) to blue for long waiting times (i.e., equilibrium initial states). Values of $t_\mathrm{w}$ normalized by $\tau_\alpha$ at $T_0$ are indicated for the highest and lowest waiting times.}
	\label{fig:Results neq}
\end{figure*}

To prepare out-of-equilibrium initial states, we use the temperature-step protocol illustrated in Fig.~\ref{fig:Method}c. The sample is first brought to temperature $T_1$, where it reaches equilibrium (i.e., $T_\mathrm{f}=T_1$), and we then apply a down step to $T_0$. Since $t_\mathrm{down}$ is large compared to the initial relaxation time $\tau(T_1)$, this down step is not ideal, which means the system ages during the cooling phase. Even so, we see that the liquid ultimately falls out of equilibrium (see. Fig.~\ref{fig:Method}c), with $T_\mathrm{f}-T_0\approx 0.5 \Delta T$ when the sample reaches $T_0$. At this stage, the temperature has decreased enough to significantly slow down the aging process and the full re-equilibration time $t_\mathrm{eq}$ of the system, which is on the order of $10 \tau_\alpha$, takes between 3~h at $T_0 = 196$~K to 30~h at 193~K. We do not wait for this full equilibration and we instead perform an ideal temperature up step back to $T_1$ after a waiting time $t_\mathrm{w}$. By repeating this experiment with different waiting times, we are able, as illustrated in Fig.~\ref{fig:Method}d, to explore the re-equilibration dynamics for initial states either equilibrated or increasingly distant from equilibrium. This protocol was conducted for three different combinations of $T_0$ and $\Delta T = T_1 - T_0$ with $t_\mathrm{w}$ between 8 min and 16 days. Results for all parameter sets are shown in Fig.~\ref{fig:Results neq}a-c.

\begin{figure*}
	\centering
	\includegraphics[width=\textwidth]{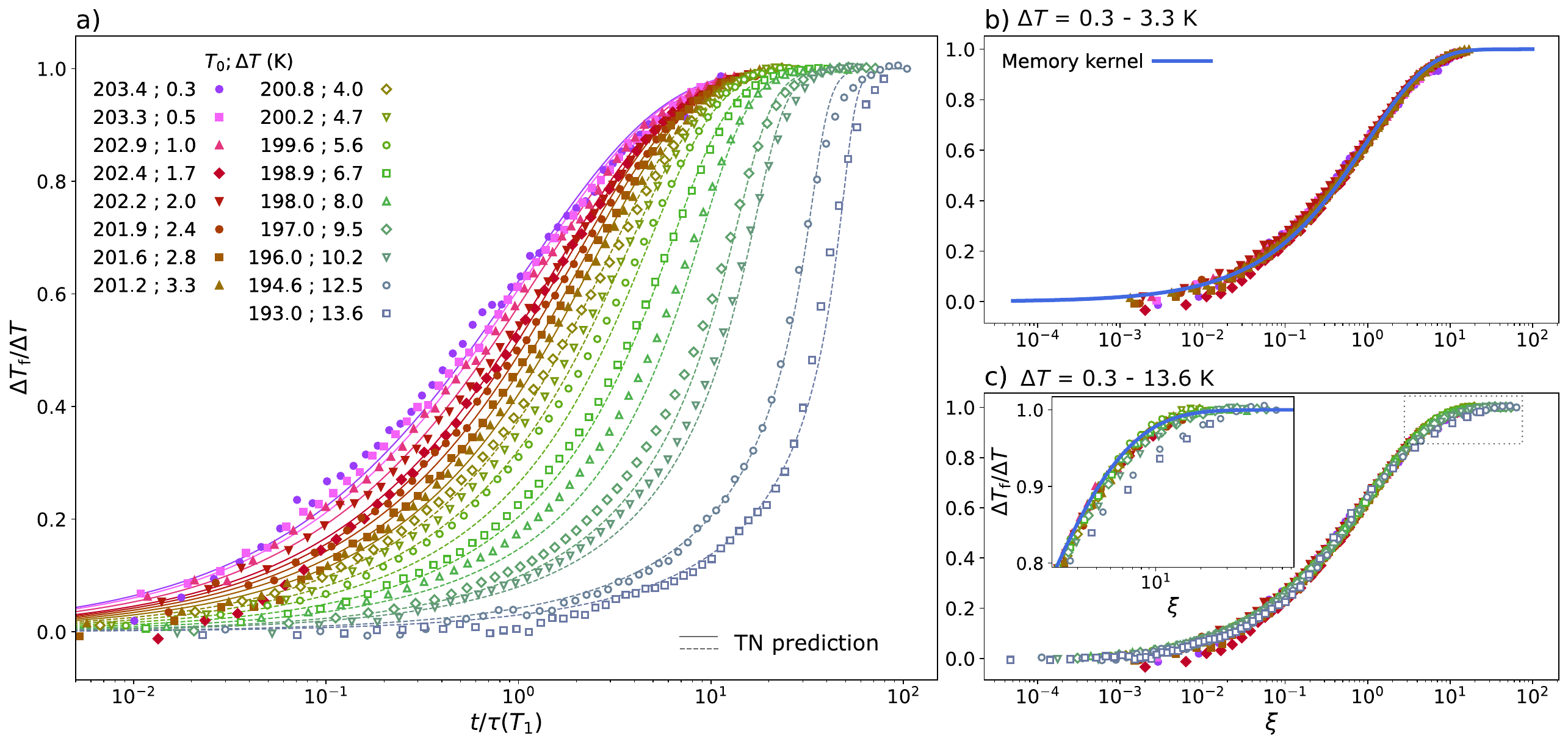}
	\caption{Normalized relaxation curves for up steps from equilibrium states of increasing amplitude, including up steps with largest waiting time $t_w$ from Fig.~\ref{fig:Results neq}, versus (a) normalized experimental time $\nicefrac{t}{\tau(T_1)}$ and (b)-(c) material time $\xi$. The small amplitude steps ($\Delta T\leq3.3$~K, shown with solid markers) were used to determine the TN parameters $x=0.36$ and $M(\xi)$ (shown with a blue line in (b)). The TN predictions are shown with solid and dashed lines.} 
\label{fig:Results eq}
\end{figure*}

\subsection{TN parameters determination}\label{subsec:level3b}
Our goal is to compare our observations to the prediction of the TN formalism. In this framework, it is assumed that the system responds linearly to any change in temperature provided that its evolution is followed as a function of a \textit{material time} $\xi$~\cite{Narayanaswamy1971}:
\begin{equation}
	T_\mathrm{f} (\xi) = T (\xi) - \int_{0}^{\xi} M(\xi-\xi') \frac{\mathrm{d}T}{\mathrm{d}\xi'}\mathrm{d}\xi'
	\label{eq:TN prediction}
\end{equation}
 Material time $\xi$ and laboratory time $t$ are linked through the out-of-equilibrium relaxation time $\tau = \mathrm{d}t/\mathrm{d}\xi$, which is itself dependent on the state of the system, described by both $T$ and $T_\mathrm{f}$~\cite{Tool1946}.

To get TN predictions, we need the memory kernel $M(\xi)$, as well as the out-of-equilibrium relaxation time $\tau(T, T_\mathrm{f})$, which is also the most delicate part to account for. Beyond the $T_\mathrm{f}=T$ case, for which we expect to recover the equilibrium relaxation time $\tau_\alpha(T)$, the effect of the distance from equilibrium on $\tau$ is unknown \textit{a priori}. One approach, originally applied in strong glass-formers~\cite{Tool1946, Narayanaswamy1971}, is to assume that:
\begin{equation}
	\tau = \tau_\alpha(T)^x \tau_\alpha(T_\mathrm{f})^{1-x}
	\label{eq:relaxation time (x)}
\end{equation}
where $x$ is the system-dependent \textit{non-linearity parameter}.

In order to determine $M(\xi)$ and $x$ for TEAC, we performed temperature up steps with amplitudes ranging from 0.3 to 3.3~K, starting from equilibrium states at temperatures between 201 and 204~K (see Fig.~\ref{fig:Results eq}a). On this temperature range, $\tau_\alpha(T)$ was evaluated through dielectric spectroscopy (see Supporting Information) and fitted to a Vogel-Fulcher-Tammann (VFT) law $\tau_\alpha(T)=\tau_\mathrm{VFT} e^{\nicefrac{A}{(T-T_\mathrm{k})}}$ (see Fig.~\ref{fig:tau_alpha}). Given the small heating time $t_\mathrm{up}$, these temperature steps can be considered ideal (\textit{i.e.} $t_\mathrm{up}\ll \tau(T=T_1, T_\mathrm{f}=T_0)$) and Eq.~\ref{eq:TN prediction} can be simplified into:
\begin{equation}
	\frac{T_\mathrm{f}(\xi)-T_0}{\Delta T} = 1-M(\xi)
	\label{eq:simplified TN prediction}
\end{equation}
This means that all re-equilibration curves corresponding to different amplitudes can be collapsed when plotted as a function of $\xi$, and that the resulting curve directly corresponds to $1-M(\xi)$. We determined $x=0.36$ to be the optimal value, leading to the collapse shown in Fig.~\ref{fig:Results eq}b, on which $M$ was fitted as a stretched exponential $M(\xi) = e^{-\left( \nicefrac{\xi}{\tau_0} \right)^\beta}$ with parameters $\tau_0=0.98$ and $\beta=0.58$.

To predict the re-equilibration dynamics starting from an out-of-equilibrium state, we cannot use Eq.~\ref{eq:simplified TN prediction}. This because we have to take into account the whole thermal history $(T(t), T_\mathrm{f}(t))$ of the sample since its last equilibrated state. For this, we use a discretized version of Eq.~\ref{eq:TN prediction} to compute the evolution of $T_\mathrm{f}(t)$ for the combination of temperature down and up steps shown in Fig.~\ref{fig:Method}c.
\begin{equation}
	T_\mathrm{f,\,i+1} = T_\mathrm{i+1} - \sum_\mathrm{j \leq i} M(\xi_{\mathrm{i}+1}-\xi_\mathrm{j})\frac{\mathrm{d}T}{\mathrm{d}\xi_\mathrm{j}}\Delta\xi_\mathrm{j}
	\label{eq:discrete TN}
\end{equation}

\subsection{$\tau_\alpha$ extrapolation on low-T}\label{subsec:level3c}
A further difficulty that needs to be addressed in order to make a TN prediction is the fact that the protocol of Fig.~\ref{fig:Method}c involves temperature as low as 193~K, which is significantly out of the range 202-225~K in which $\tau_\alpha(T)$ could be measured with dielectric spectroscopy. Such a distant extrapolation of the VFT law would be too sensitive to the uncertainties on its parameters and would likely lead to unreliable predictions.

\begin{figure}
	\centering
	\includegraphics[width=\columnwidth]{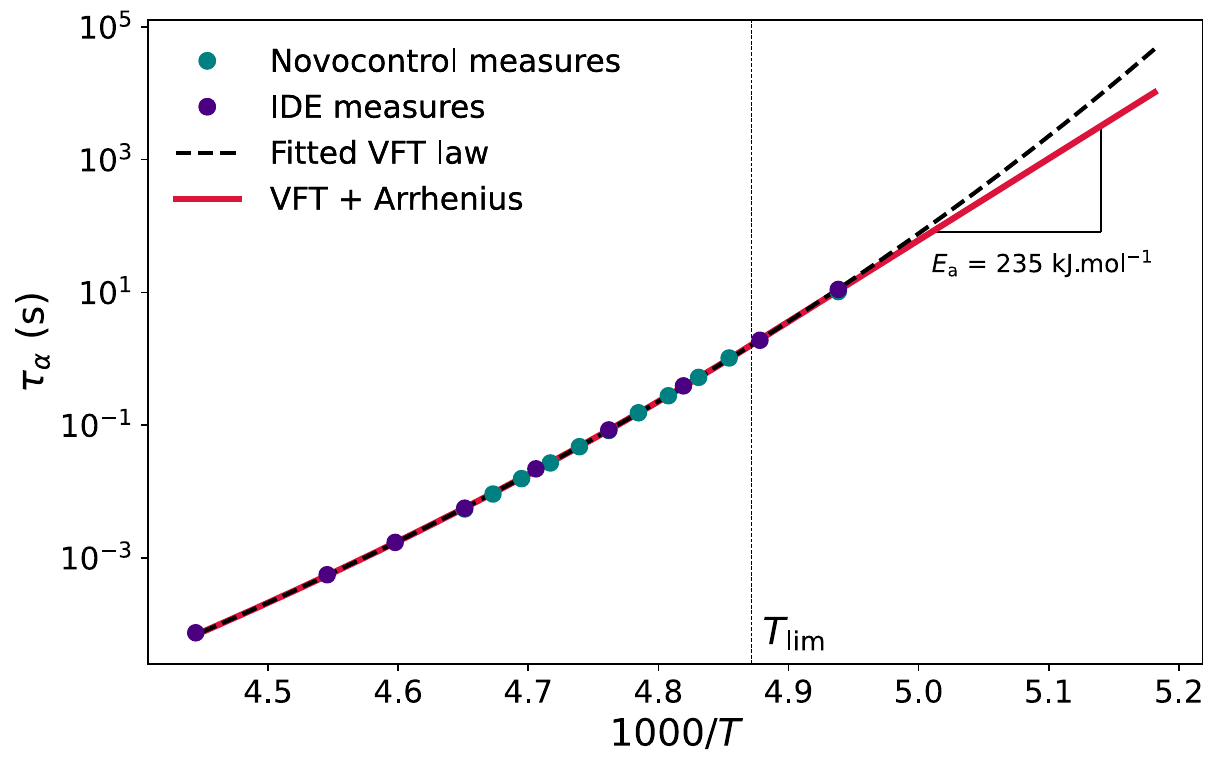}
	\caption{Equilibrium relaxation time $\tau_\alpha$ for TEAC measured by dielectric spectroscopy using a Novocontrol setup (blue dots) and with our own experimental setup (purple dots). The black dashed line corresponds to the VFT fit to these data ($\tau_0 = 1.24\cdot10^{-18}$~s, $A = 2.59\cdot10^{3}$~K$^{-1}$, $T_\mathrm{k} = 143$~K). The red curve is the one we used in the rest of this article. It corresponds to the VFT fit for $T>T_\mathrm{lim} = 205.3$~K, and below to an Arrhenius law $\tau_\alpha(T) = \tau_\mathrm{A} e^{\nicefrac{E_\mathrm{a}}{RT}}$with $\tau_\mathrm{A}=2\cdot10^{-60}$~s and an activation energy $E_\mathrm{a} = 235$~kJ$\cdot$mol$^{-1}$.}
	\label{fig:tau_alpha}
\end{figure}

To overcome this limitation, we chose to constrain $\tau_\alpha(T)$ at low temperature using the aging experiments following the temperature down steps. The experimental data for the three datasets are shown in Fig.~\ref{fig:Comparison data-TN downstep} with markers and solid lines. We used the TN parameters ($M$ and $x$) determined above and we parameterized $\tau_\alpha(T)$ for $T<T_\mathrm{lim}$ by assuming an Arrhenius dependence with an activation energy $E_\mathrm{a}$ and by requiring continuity of the derivative (see Fig.~\ref{fig:tau_alpha}). For each dataset, we computed $T_f(t)$ using Eq.~\ref{eq:discrete TN} with the measured $T(t)$ as input and we optimized the value of $T_\mathrm{lim}$, which in turn constrain $E_\mathrm{a}$ to obtain the best overall agreement. The result for $E_\mathrm{a}= 235$~kJ.mol$^{-1}$ and $T_\mathrm{lim}=205.3$~K is shown as dotted lines in Fig.~\ref{fig:Comparison data-TN downstep} and leads to a good agreement, especially for $T_0 = 196$~K and 194.6~K.

\subsection{TN predictions}\label{subsec:TNprediction}
\begin{figure}
	\centering
	\includegraphics[width=\columnwidth]{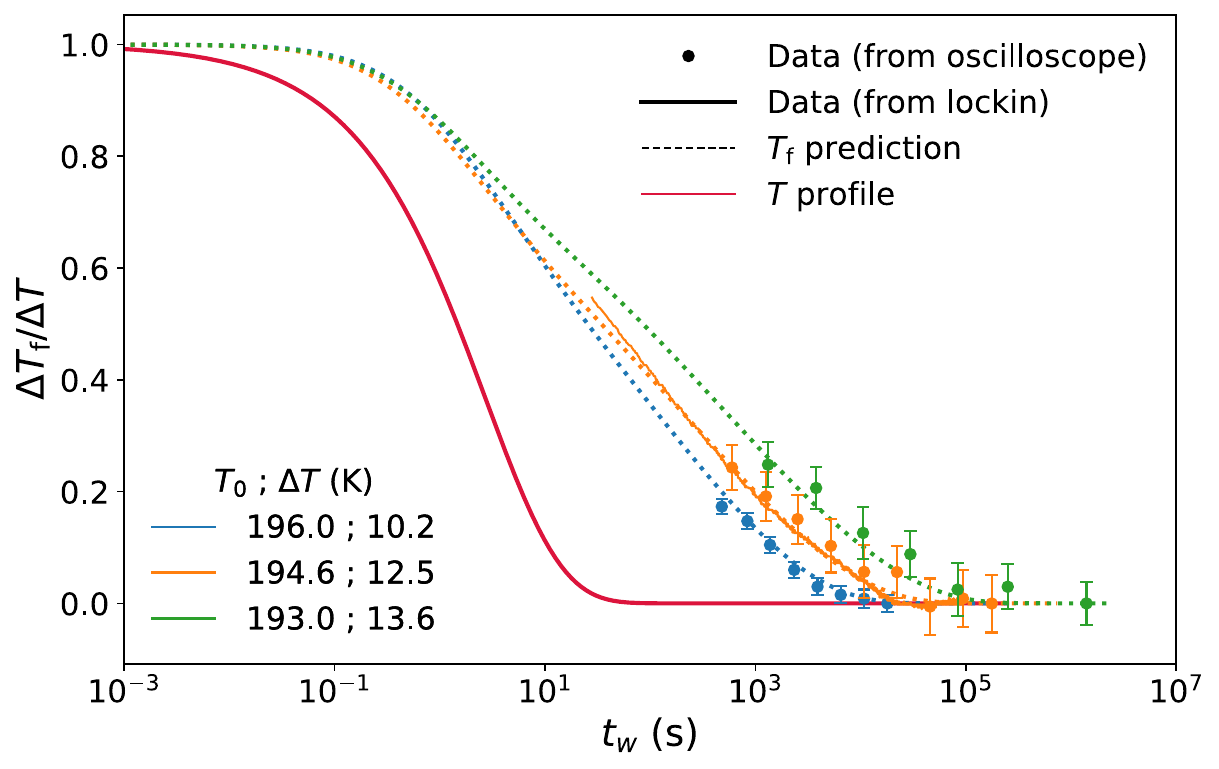}
	\caption{Normalized relaxation curves for the the down step experiments. The temperature profile $T(t)$ in shown in red and the experimental data are shown with markers and with a solid line for $T_0,\,\Delta T = 194.6,\,12.5$~K. The TN prediction from Eq.~\ref{eq:discrete TN} are shown with dotted lines.}
	\label{fig:Comparison data-TN downstep}
\end{figure}

\begin{figure}
	\centering
	\includegraphics[width=\columnwidth]{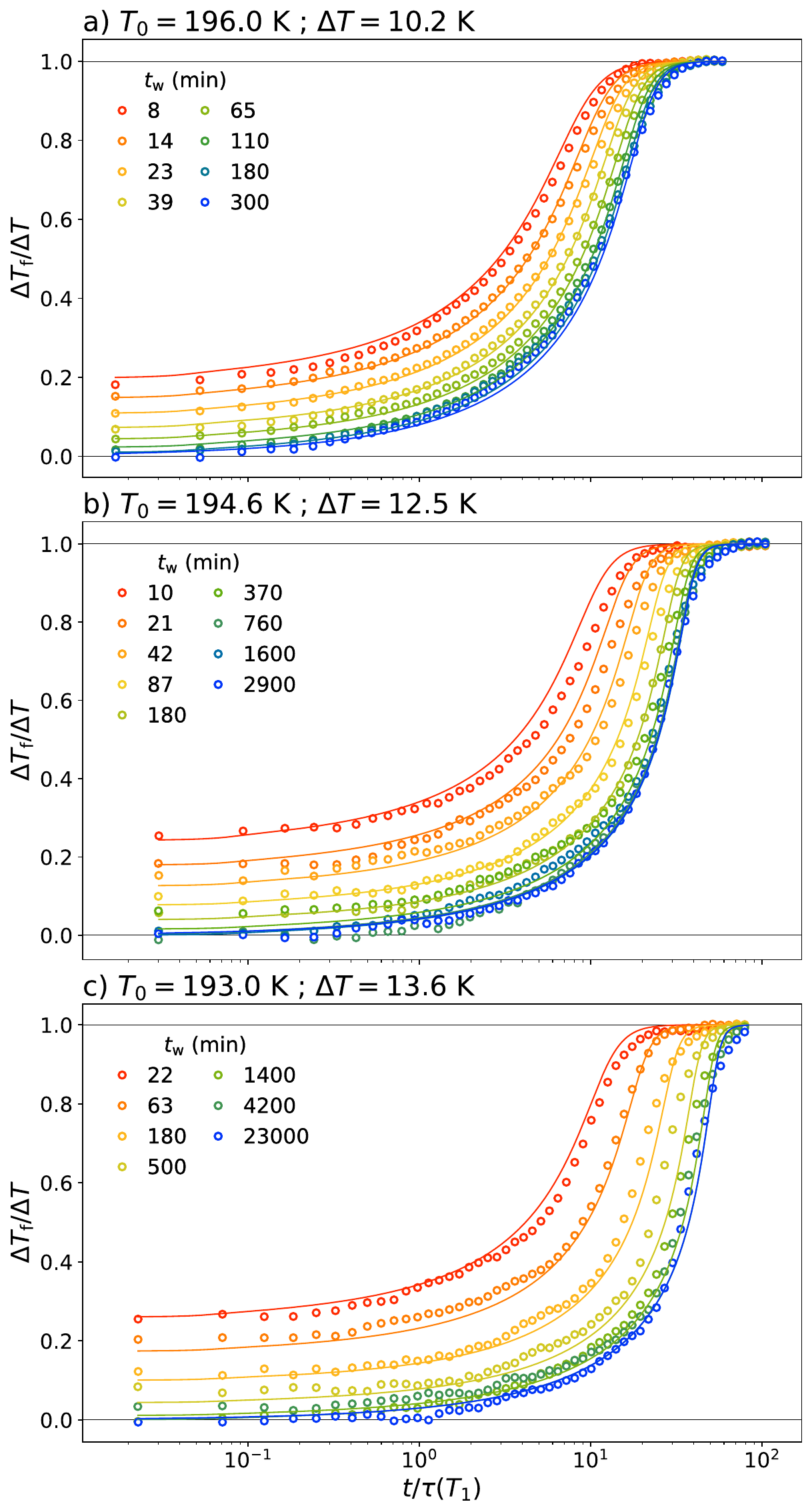}
	\caption{Normalized relaxation curves for up steps from out-of-equilibrium states for three pairs of $T_0$, $\Delta T$ (a,b,c). The TN predictions calculated from Eq.~\ref{eq:discrete TN} are shown with solid lines.}
	\label{fig:Results neq + pred}
\end{figure}

We now have all the necessary ingredients to make TN predictions for each aging experiment. The results are shown in Fig.~\ref{fig:Results eq}a as solid lines for steps involved in the determination of the TN model parameters ($\Delta T = 0.3 - 3.3$~K) and as dashed lines for the larger steps ($\Delta T = 4.0 - 13.6$~K). The predictions for the steps starting from out-of-equilibrium steps are obtained using Eq.~\ref{eq:discrete TN} with the $T(t)$ profile shown in Fig.~\ref{fig:Comparison data-TN downstep} as input. The resulting evolution of $T_\mathrm{f}(t)$ covers the down steps (see dashed lines in Fig.~\ref{fig:Comparison data-TN downstep}) and the up steps (see solid lines in Fig.~\ref{fig:Results neq + pred}).

\begin{figure*}
	\centering
	\includegraphics[width=\textwidth]{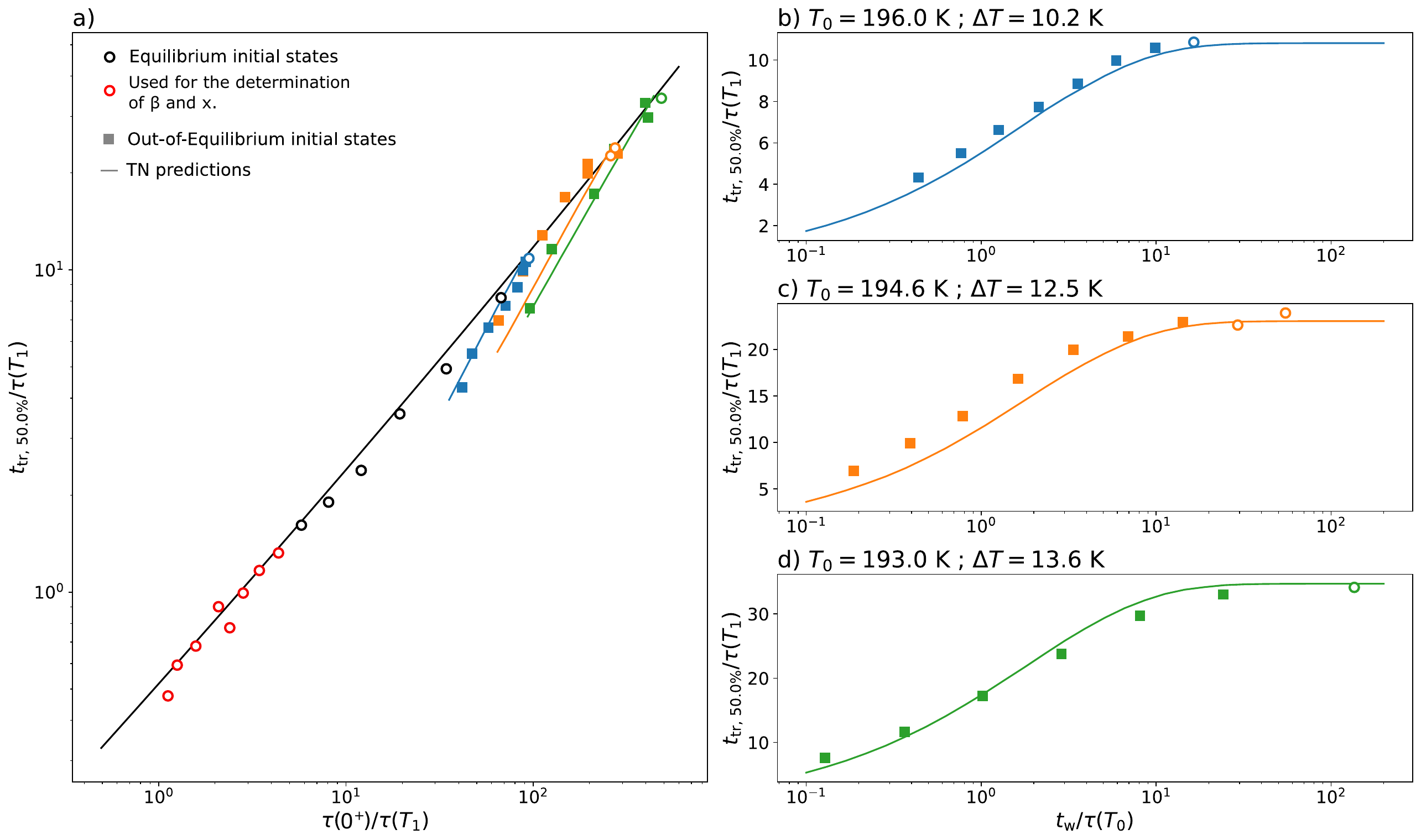}
	\caption{(a) Transformation time at 50~\% between $T_\mathrm{f}(0)$ and $T_1$, normalized by the final relaxation time $\tau_\alpha (T_1)$ as a function of the contrast $\tau(0^+)/\tau(T_1)$ between the initial out-of-equilibrium relaxation time and final relaxation time. Empty circles correspond to up steps from equilibrium states, while solid squares start from increasingly out-of-equilibrium initial states. Solid lines corresponds to TN prediction. (b-d) Transformation time as a function of the waiting time at $T_0$ normalized by the equilibrium relaxation time for three pairs of $T_0$, $\Delta T$.}
	\label{fig:ttr}
\end{figure*}

\section{Discussion}\label{sec:level4}

\subsection{Low-T extrapolation on the $\tau_\alpha(T)$ function}\label{subsec:level4a}
Before discussing the results, let us first address the fact that in order to obtain an accurate TN predictions for the temperature down steps of Fig.~\ref{fig:Comparison data-TN downstep}, we had to rely on a $\tau_\alpha(T)$ curve which substantially deviates near and below $T_\mathrm{g}$ from the extrapolation of the VFT law (see Fig.~\ref{fig:tau_alpha}). This is not a complete surprise as it has long been realized that such a VFT extrapolation overestimate the low temperature relaxation time for many molecular glass-formers~\cite{Stickel1996, hecksher2008little, Novikov2015}. The present experiment constitutes an indirect measurement of the equilibrium relaxation time below $T_\mathrm{g}$ from an aging experiment. Although this conclusion should be considered with caution, as it assumes that the TN approach holds on down steps below $T_\mathrm{g}$, aging experiments performed on polymeric systems below $T_\mathrm{g}$ have also reported an overestimation of the relaxation times from a VFT extrapolation~\cite{zhao2013using,yoon2018testing}.

We would like to emphasize the fact that we chose an Arrhenius law as a practical way of parameterizing $\tau_\alpha(T)$ at low temperature with only one parameter. We expect a second VFT with different coefficients~\cite{Stickel1996}, a parabolic or a more complex law~\cite{Novikov2015} to work just as well. We note in passing that although the various competing theories of the glass transition have made predictions for the functional dependence of $\tau_\alpha(T)$~\cite{Bouchaud2004, Tarjus2005, Elmatad2010}, it remains impracticable to distinguish theories based on this alone. For instance, the VFT law can be directly obtained form the RFOT theory~\cite{Bouchaud2004,lubchenko2004theory}, but taking into account facilitation~\cite{Scalliet2022} could affect these results, especially at low temperature as advocated in Ref.~\cite{lubchenko2004theory}.

\subsection{Equilibrium initial states}\label{subsec:level4b}
We now turn to the comparison between experimental data and the TN prediction, starting from equilibrated initial states. As visible in Fig.~\ref{fig:Results eq}a and b, the agreement is excellent for $\Delta T\leq 3.3$~K. This is expected for temperature steps of small to moderate amplitude~\cite{Riechers2022} and from the fact that the TN parameters were fitted on these data. If we consider the steps of larger amplitude, we observe a generally decent agreement, especially concerning the prediction of the timescale at which the equilibration occurs. This can be characterized quantitatively through the transformation time at 50~\% $t_\mathrm{tr,\,50\%}$, which is the time at which a relaxation curve has completed half of its trajectory between its initial ($T_\mathrm{f}(0)=T_0)$ and final state ($T_1$). This quantity is shown in Fig.~\ref{fig:ttr}a with empty markers as a function of the ratio between the initial out-of-equilibrium relaxation time $\tau(0^+)=\tau(T=T_1, T_\mathrm{f}=T_0)$ and the final relaxation time $\tau_1$. This quantity was chosen because it characterizes the contrast in relaxation time between the initial state just after the temperature step and the equilibrium state that the system will eventually reach. The TN prediction, shown as a black line, matches very well the data especially for $\Delta T\leq 3.3$~K and $\Delta T>10$~K. We attribute this unexpected success for large $\Delta T$ to the fact that it explores the low temperature range where $\tau_\alpha(T)$ was optimized on aging data. Yet, this was done on down steps and while the corresponding up steps share the same initial and final temperatures, the path taken by the sample through the ($T$, $T_\mathrm{f}$) space are entirely different. The fact that it leads to such a good prediction for the up steps is thus a sign of the robustness of the TN.

\begin{figure}
	\centering
	\includegraphics[width=0.8\columnwidth]{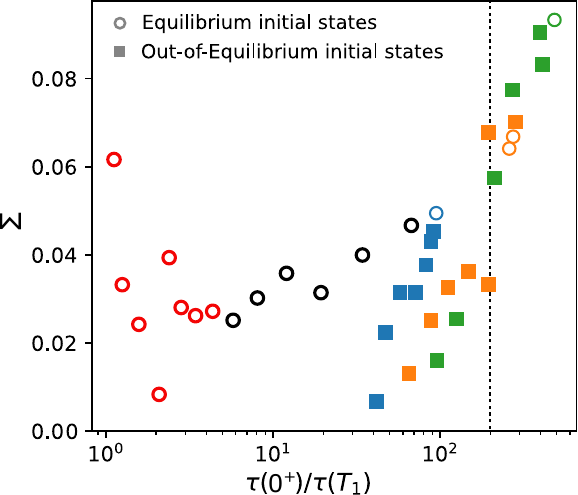}
	\caption{Maximum deviation $\Sigma$ between the experimental data and the TN prediction horizontally shifted to match $t_\mathrm{tr,\,50\%}$, as a function of the contrast $\tau(0^+)/\tau(T_1)$. The colors are the same as in Fig~\ref{fig:ttr}. The vertical dashed line situate the limit of validity of the TN.}
	\label{fig:Sigma}
\end{figure}

If we consider not only the transformation time but also the shape of the re-equilibration curve, we clearly notice in Fig.~\ref{fig:Results eq}a a discrepancy between the data and the prediction at the end of the equilibration dynamics for the largest temperature steps. This can also be seen when attempting to collapse all curves as a function of the material time (see Fig.~\ref{fig:Results eq}c). Large $\Delta T$ experiments fall systematically below $M(\xi)$ at $\xi \approx 10$ (see inset of Fig.~\ref{fig:Results eq}c). We would like to characterize this discrepancy on the curve shape independently of that of $t_\mathrm{tr,\,50\%}$. Indeed, the latter can be corrected by slightly modifying $\tau_\alpha(T)$, whereas the former is a more profound sign of the limit of the material time approach. To do that, we horizontally shift the TN prediction for $\Delta T_\mathrm{f}(t)/\Delta T$ so that it matches perfectly the data at 50~\% (\textit{i.e.} we force $t_\mathrm{tr,\,50\%}$ to be correct) and we define $\Sigma$ as the maximum difference between the prediction and the data for $\Delta T_\mathrm{f}(t)/\Delta T$. This quantity is shown in Fig.~\ref{fig:Sigma} with empty circles as the function of the contrast $\tau(T_1, T_0)/\tau(T_1)$. We observe that it becomes significant ($>5~\%$) only for the three largest steps, \textit{i.e.} for $\Delta T>10$~K. Here, it seems that for TEAC, we reach the limit of validity of the TN in the sense that even if $t_\mathrm{tr,\,50\%}$ is well predicted, the equilibration dynamics cannot be mapped to the linear response through a unique material time. This might be a sign that the microscopic dynamics at such a distance from equilibrium starts to differ from equilibrium dynamics. 

\subsection{Out-of-equilibrium initial states}\label{subsec:level4c}
Let us now discuss the case of temperature steps starting from out-of equilibrium states $T_\mathrm{f}(0)>T_0$, shown in Fig.~\ref{fig:Results neq + pred}. It can already be noted that shorter waiting times lead to faster relaxation (see Fig.~\ref{fig:Method}d and Fig.~\ref{fig:Results neq}). This behavior can be understood in the TN framework from the fact that these initial out-of-equilibrium states originate from the equilibrium at $T_1$: the less time they are given to equilibrate at $T_0$, the more they remember their previous state through memory effects, and thus the faster they return to it. For $T_0 = 196$~K and $\Delta T = 10.2$~K, the agreement with TN is excellent, both concerning the transformation time (see Fig.~\ref{fig:ttr}a-b) and the shape of the re-equilibration curves, despite these data not being used to determine the TN parameters. This test of the TN is analogous to the $-2$~K/$+1$~K double jump experiments reported in ref.~\cite{Riechers2022} although in the present case, the distance from equilibrium reaches $-4$~K and $+10$~K. It should be noted that the good fit of the data in Fig.~\ref{fig:Comparison data-TN downstep} likely contributes to this success.

 For larger steps, corresponding to $T_0=196$~K and 193~K, the agreement is not as good. Yet, the evolution of the transformation time as a function of the waiting time is qualitatively well reproduced (see Fig.~\ref{fig:ttr}c-d). The systematic overestimation at 196~K could result from the inaccuracy of $\tau(T)$ on the relevant temperature range. The quality of the prediction regarding the full re-equilibration dynamics, characterized by $\Sigma$, is shown in Fig.~\ref{fig:Sigma} with full squares. When plotted as a function of $\tau(0^+)/\tau_1$, they appear consistent with the steps starting from equilibrium: For $\tau(0^+)/\tau_1<100$, the maximum deviation remains smaller than 5~\%, which is good given the experimental uncertainties, before gradually deteriorating. Looking at the square markers in Fig.~\mbox{\ref{fig:Sigma}}, it appears that $\Delta T_\mathrm{f}$, rather than the temperature step amplitude $\Delta T$, is the more relevant parameter. Indeed, while the TN model does not perform very well for the sample equilibrium for 16 days ($\Delta T_\mathrm{f} = \Delta T = 13.6~K$, green circle), it gives a good prediction for those annealed for 63~min or less corresponding to $\Delta T_\mathrm{f} \approx 10$~K (the two leftmost green squares). This means the contrast between the initial and final states matters more than the jump amplitude in Kelvins.

\subsection{TN formalism vs single parameter aging ansatz}
\begin{figure*}
	\centering
	\includegraphics[width=\textwidth]{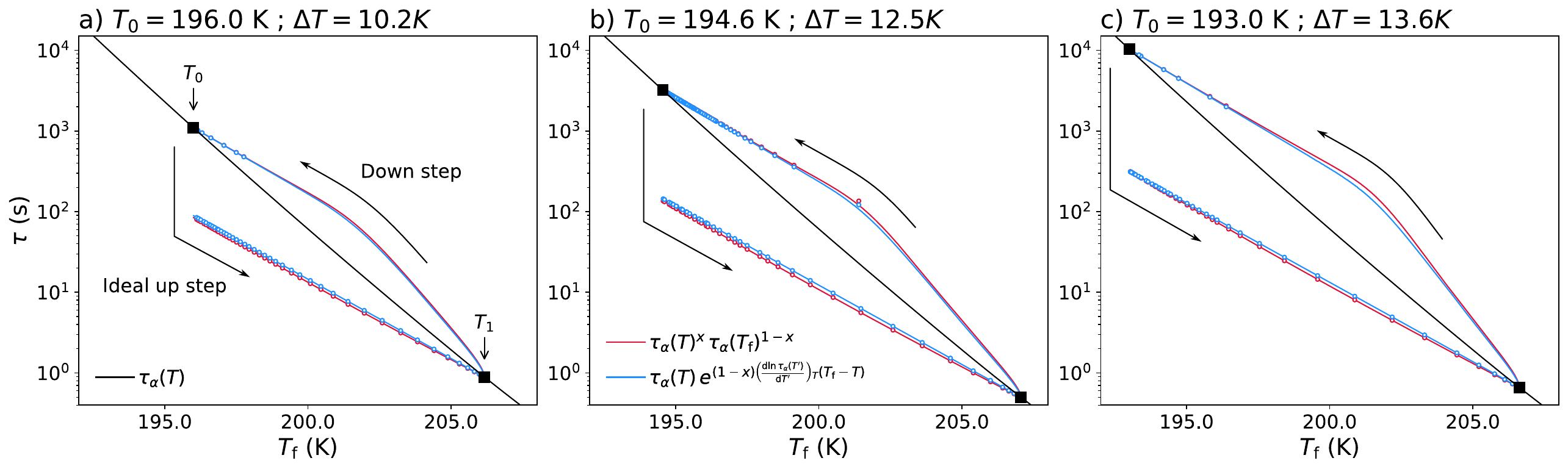}
	\caption{(a-c) Out-of-equilibrium relaxation time $\tau(T, T_\mathrm{f})$ for the three double temperature steps reaching equilibrium at $T_0$. The red curves and markers correspond to the quantity $\tau$ computed in the framework of the TN (Eq.~\ref{eq:relaxation time (x)}) while its Taylor expansion used in the SPA Ansatz (Eq.~\ref{eq:relaxation_time_SPA)}) is shown in blue.}
	\label{fig:TN_SPA}
\end{figure*}

The single parameter aging (SPA) ansatz is a slightly simplified version of the TN formalism that was shown to give an excellent agreement with aging experiments on molecular liquids~\cite{Hecksher2010, Hecksher2015, Roed2019, Hecksher2024}. It is particularly suited in the case of ideal temperature steps, for which it gives a straightforward way of predicting the aging following a temperature step from the data obtained for a temperature step of different amplitude and a unique material dependent parameter~\cite{Hecksher2015}. The differences between the SPA ansatz and the TN approach can be summarized in three points: (i) the observable used to monitor the aging dynamics is employed directly, without conversion to a fictive temperature. (ii) the memory kernel is extracted directly from the experimental data rather than being fitted with a stretched exponential. (iii) the out-of-equilibrium relaxation time is obtained through a Taylor expansion of Eq.~\ref{eq:relaxation time (x)} around the equilibrium temperature. Expressed in the formalism of this work, it leads to:
\begin{equation}
	\tau = \tau(T_1) \exp{\left[(1-x)\left.\frac{\mathrm{d}\ln \tau}{\mathrm{d}T}\right|_{T_1} (T_\mathrm{f}-T_1)\right]}
	\label{eq:relaxation_time_SPA)}
\end{equation}
While we do not expect points (i) and (ii) to affect the prediction significantly, point (iii) can, in principle, have a strong effect, especially for large steps. In order to check whether it is the case here, we computed $\tau(T, T_\mathrm{f})$ for the three double steps ($\Delta T = 10.2$, 12.5, 13.6~K) that reach equilibrium at $T_0$ using both expressions (Eqs.~\ref{eq:relaxation time (x)} and \ref{eq:relaxation_time_SPA)}) with the results shown in Fig.~\ref{fig:TN_SPA}. The two quantities differ only slightly, indicating that the SPA ansatz provides aging predictions of comparable quality to those of the TN approach (see supplements). This can be attributed to the fact that, for TEAC, over the relevant temperature range $\ln \tau_\alpha(T)$ is almost linear with $T$. This would not have been the case if we had used an extrapolation of the VFT law for $\tau_\alpha(T)$. But, in the present case, $\tau_\alpha(T)$ below $T_\mathrm{g}$ is very much constrained by down step aging experiments and although its precise form is imposed, we believe that the near-linearity of $\ln \tau_\alpha(T)$ with $T$ is robust. Numerous forms of $\tau(T, T_\mathrm{f})$ have been proposed in the literature~\cite{mauro2009nonequilibrium}, with Eq.~\ref{eq:relaxation time (x)} representing only one possible choice. This result suggests that the simplest expression, Eq.~\ref{eq:relaxation_time_SPA)}, is sufficient and require no further refinement, consistently with a previous study on glycerol~\cite{Henot2024}.

\subsection{A threshold for the limit of validity?}\label{subsec:threshold}
Given that the TN gives excellent prediction for small to moderate temperature steps~\cite{Riechers2022} but that it cannot account for the heterogeneous re-equilibration evidenced for very large temperature steps, it would be interesting to precisely identify its limits. However, this faces several limitations: (i) TN can be applied to a diversity of systems: from silica glasses~\cite{lancelotti2024kinetics} to amorphous polymers~\cite{Hodge1994} having various fragilities and non-exponentiality coefficients~\cite{malek2024distinguish}. This already limits \textit{a priori} our ability to define a simple general threshold. Even if we restrict ourselves to organic glass-formers for which the non-exponentially might not be too far from $\beta=0.5$~\cite{pabst2021generic}, a general criterion could not be expressed simply as a maximum step amplitude $\Delta T$. In the present work we chose to characterize the step amplitude in Fig.~\ref{fig:ttr} and \ref{fig:Sigma} through the initial contrast $\tau(0^+)/\tau_1$ which can be seen as the driving force of the re-equilibration. The transformation time $t_\mathrm{tr,\,50\%}$~\cite{VilaCosta2023} or the stabilization period~\cite{malek2024distinguish}, could also be used with the advantage that they are an outcome of the phenomenon and can thus directly be obtained from the experimental data independently of any model. (ii) The TN approach likely fails in a continuous manner~\cite{Henot2024, amari2026large} and such a threshold may depends on the accessible observable and its uncertainty. (iii) The quality of a TN prediction depends on the parameters of the model. While for strong glass-formers they reduce to only two scalars~\cite{malek2024distinguish}, it is no longer the case for a fragile glass-former for which the precise knowledge of $\tau_\alpha(T)$ is required, precisely in the sub-$T_\mathrm{g}$ range where it is challenging to access. It was already shown in the supplementary of ref.~\cite{Riechers2022}, that predictions for large up steps on NMEC improved significantly when the clock rate was taken from fits of the aging curves themselves rather than from the VFT extrapolation of $\tau_\alpha(T)$. Here, we show that letting some freedom on $\tau_\alpha(T)$ allows one to get a good agreement with the TN for up to $\Delta T_\mathrm{f} = 10$~K. For larger steps, we observe noticeable difference on the relaxation dynamics that cannot be suppressed by adjusting the relaxation time.

We identify the limit of validity for the TN for TEAC for temperature up steps associated with an initial contrast of $\tau(0^+)/\tau(T_1) \approx 200$ (see Fig.~\ref{fig:Sigma}). This corresponds to a transformation time $t_\mathrm{tr~50\%}/\tau(T_1)\approx 10$. This is consistent with a previous study of glycerol with the same experimental setup were noticeable deviations to the material time approach appeared for up steps of 15-18~K. The TEAC being 15~\% more fragile than glycerol, we expect the $\Delta T$ limit to be higher for glycerol. It does seem to be the case as these steps correspond for glycerol to $t_\mathrm{tr~50\%}/\tau(T_1)= 7-12$ or $\tau(0^+)/\tau(T_1) = 80-140$. On a model binary Lenard-Jones system, Amari~\textit{et al.}~\cite{amari2026large} were able to test very precisely the existence of a material time as well as its ability to collapse different observables following temperature up steps. They observed a very good collapse for an up step corresponding to $t_\mathrm{tr~50\%}/\tau(T_1)\approx 5$ but a much poorer collapse for an up step with $t_\mathrm{tr~50\%}/\tau(T_1)\approx 20$. This appears qualitatively consistent with the present result although the method and the system are extremely different. By performing calorimetric experiments on a molecular glass prepared either from the liquid phase or from vapor deposition, Vila-Costa~\textit{et al.}~\cite{VilaCosta2023} identified the limit between homogeneous and heterogeneous equilibration at $t_\mathrm{tr}/\tau(T_1) = 180 \pm 60$ (this is a full transformation time, larger than $t_\mathrm{tr~50\%}$ by a factor 2-3). This again is qualitatively consistent with the present findings as we expect that when increasing the step amplitude, the inability of the TN to account for the experimental data with an accuracy of a few percent should precede the case of a clearly heterogeneous re-equilibration. If we extrapolate the trend of Fig.~\ref{fig:ttr}, this criterion would correspond for TEAC to $\tau(0^+)/\tau(T_1) \approx 1000$ and thus to $\Delta T_\mathrm{f} \approx 20 \pm 5$~K. Although this is only a rough estimate relying on an extrapolation of the TN model out of its validity range and on a generalization between different liquids, it shows that the heterogeneous re-equilibration regime may not be out of reach, even for an ideal temperature step experiment starting from equilibrium. 

\section{Conclusion}\label{sec:level5}
In this work, we measured the re-equilibration dynamics of a liquid following temperature up steps starting either from equilibrium or out-of-equilibrium. The latter case was realized through double step experiments leading to well characterized initial states. We tested the TN model, which eventually proved to be equivalent to the SPA ansatz in the regime we explored, by determining its parameters for moderate amplitude up steps ($\Delta T \leq 3.3$~K) and by estimating the equilibrium relaxation time below $T_\mathrm{g}$ from temperature down steps in a self-consistent manner. We then compared the experimental data with TN predictions and found a surprisingly good agreement regarding the transformation time at 50~\%, \textit{i.e.} the time scale of re-equilibration of the system, even for steps as large as 13.6~K. The shape of the re-equibration curve, however, started to show noticeable discrepancies with TN predictions for steps of $\Delta T_\mathrm{f}>10$~K. This corresponds to a contrast between the initial and final relaxation time $\tau(0^+)/\tau_1$ of $\approx 10^2$ and a transformation time $t_\mathrm{tr,\,50\%}/\tau_1 \approx 10$. This threshold for the limit of validity of the TN appears consistent with previous work from the literature on others systems~\cite{VilaCosta2023, Henot2024, amari2026large}.

The discrepancy we observe for large steps however remains quite small and the regime studied here still does not correspond to the clear heterogeneous re-equilibration observed in ultra-stable glasses. Performing significantly large up step experiments starting from equilibrium remains challenging as it needs to be either extremely fast or to start from a temperature at which full equilibration is prohibitively long. Yet, as discussed above, the heterogeneous re-equilibration might not be out of reach. In the future, we hope to take advantage of the possibility offered by our experimental setup to perform fast temperature up step while monitoring the dielectric response to explore this regime.

\section*{Supplementary Material}
The supplementary material provides details on the determination of the fictive temperature and the equilibrium relaxation time and on the test of the SPA ansatz.

\begin{acknowledgments}
	The author thanks J. Dyre, K. Niss, M. Gonzalez Silveira and C. Raepsaet for fruitful discussions and for providing feedback on the manuscript.
\end{acknowledgments}

\section*{Data availability statement}
The data that support the findings of this study are available from the corresponding author upon reasonable request.

\providecommand{\noopsort}[1]{}\providecommand{\singleletter}[1]{#1}%

\end{document}